\documentclass[10pt,a4paper,english,aps,preprint]{revtex4}
\usepackage[T1]{fontenc}
\usepackage[latin9]{inputenc}
\usepackage{color}
\usepackage{textcomp}
\usepackage{amsmath}
\usepackage{amssymb}
\usepackage{esint}

\makeatletter

\@ifundefined{textcolor}{}
{%
 \definecolor{BLACK}{gray}{0}
 \definecolor{WHITE}{gray}{1}
 \definecolor{RED}{rgb}{1,0,0}
 \definecolor{GREEN}{rgb}{0,1,0}
 \definecolor{BLUE}{rgb}{0,0,1}
 \definecolor{CYAN}{cmyk}{1,0,0,0}
 \definecolor{MAGENTA}{cmyk}{0,1,0,0}
 \definecolor{YELLOW}{cmyk}{0,0,1,0}
}

\makeatother

\makeatother

\usepackage{babel}
\begin{document}

\title{Nonlinear optical refraction of the dye-doped E7 thermotropic liquid
crystal at the nematic-isotropic phase transition}

\author{V. M. Lenart}

\author{G. K. da Cruz}

\author{S. L. Gómez}

\email{sgomez@uepg.br}

\affiliation{Department of Physics, Universidade Estadual de Ponta Grossa, Ponta
Grossa, PR, Brazil.}

\author{R. F. Turchiello}

\affiliation{Department of Physics, Universidade Tecnológica Federal do Paraná,
Ponta Grossa, PR, Brazil.}

\author{I. H. Bechtold}

\affiliation{Department of Physics, Universidade Federal de Santa Catarina, Florianópolis,
SC, Brazil.}

\author{A. A. Vieira}

\author{H. Gallardo}

\affiliation{Department of Chemistry, Universidade Federal de Santa Catarina,
Florianopolis, SC, Brazil.}

\date{\today}
\begin{abstract}
It is known that the doping of liquid crystal with dyes usually changes
the physical properties of the host, like the transition temperatures
and the optical absorption among others. In this work we report a
study of the nonlinear optical refraction of a dye doped sample of
the E7 thermotropic liquid crystal by the Z-scan technique. It was
found that the nonlinear refraction of the sample is higher than the
undoped one, diverging at the clearing point. Close to the N-I transition,
the nonlinear birefringence is characterized by a critical exponent
that seams to confirm the tricritical hypothesis of the nature of
the N-I phase transition, being independent of the doping.
\end{abstract}
\maketitle

\section{Introduction}

In the field of liquid crystals, the Nematic-Isotropic (N-I) phase
transition is the most studied phenomenon, both theoretically and
experimentally, due to puzzling facts about its nature. Although initially
it was suggested that the N-I transition should be of first order\cite{Degennes-1,Longa},
new experimental data show that it can be of a tricritical character\cite{Keyes,Thoen,Syed,Chirtoc,Jadzyn,Yildiz,Erkan,Mukherjee}.
Recently we reported on the nonlinear optical properties of the E7
liquid crystal at the N-I phase transition where it was shown that
close to the clearing temperature, the nonlinear birefringence diverges
following an exponential law with an exponent that seams to support
the tricritical hypothesis\cite{Vinicius}.

The doping of liquid crystals with dyes usually leads to changes in
the optical properties of the host. There have been reported not only
changes in the optical absorption but also changes in the order parameter
and transition temperatures \cite{Bauman}, the optical Fréedericksz
transition threshold\cite{Janossy} and the nonlinear optical properties\cite{khoo,khoo2,Simoni2},
being dependent on the characteristics of the dyes. In a previous
work, the doping of the comercial liquid crystal E7 with a rod-like
fluorescent dye (4,7-Bis(2-(4-(4-decylpiperazin-1-yl)phenyl)ethynyl)-
{[}2,1,3{]}-benzothiadiazole) was investigated and the optical properties
characterized \cite{Alliprandini}. The aligned liquid crystal host
induced the ordering of the dye molecules, resulting in highly polarized
light emission.

The main purpose of this article is to report on the nonlinear optical
refraction of the E7 liquid crystal in the neighborhood of the N-I
phase transition under the addition of the {[}2,1,3{]}-benzothiadiazole
dye, employing the Z-scan technique. This report is organized as follows:
in section two we describe the samples and the Z-scan technique. In
section 3 we present the experimental results and the discussions.
Finally, in section 4, we withdraw the conclusions.

\section{Experimental details}

The abbreviation E7 stands for a liquid crystal mixture consisting
of several types of cyanobiphenyls, mainly 5CB, and in less quantity
triphenyls. It exhibits a nematic phase in the temperature interval
from $-10\,^{\circ}C$ till the transition to the isotropic phase
at $T_{NI}=58.1\,^{\circ}C$. In this work we used comercial E7 (Merck)
without further purification. The synthesis and optical properties
of the {[}2,1,3{]}-benzothiadiazole dye were published elsewhere \cite{Vieira,Alliprandini}.
The dye-doped liquid crystal samples were prepared with three different
molar concentrations of dye: 0.025\%w, 0.075\%w and 0.2\%w, labeled
S1, S2 and S3 for shortness, respectively. A fourth sample without
doping was used for comparison (labeled S0). Samples were conditioned
in parallel glass cells, separated by $20\,\mu m$ thick spacers.
The glass plates were coated with PVA (polyvinyl alcohol) and buffed
for homogeneous planar alignment of the liquid crystal. The sample
was placed in a hot stage (Instec) with a computer-controlled translational
stage (Newport). The temperature of the samples were controlled with
a precision of $0.2\,^{o}C$ between $20\,^{o}C$ and the $T_{NI}$.
The clearing temperature of the mixed system liquid crystal+dye depends
on the percentage of the dye. The $T_{NI}$ transition temperatures
of the samples S0, S1, S2 and S3 were measured by optical polarized
light microspcopy using a temperature controlled hot stage (Instec)
where the values of $58.1\,^{\circ}C$, $59.0\,^{\circ}C$, $59.4\,^{\circ}C$
and $60.1\,^{\circ}C$, respectively, were obtained.

The Z-scan technique exploits the formation of a lens in a medium
by a focused Gaussian-profile laser beam, measuring the transmitted
intensity of the sample as a function of position along the direction
of propagation of the beam (the $z$-axis)\cite{Sheik}. The magnitude
and sign of the nonlinear refraction $n_{2}$ can be obtained by means
of the closed-aperture \emph{Z}-scan technique (Fig. 1). In the closed-aperture
setup the transmitted intensity is measured behind an iris centered
along the $z$-axis. A medium characterized by a $n_{2}>0\:(n_{2}<0)$
behavies like a positive (negative) lens. The transmittance of the
sample at the \emph{z} position, normalized by the transmittance of
the sample at a far position of the focus of the lens ($z\rightarrow\infty$),
in the closed-aperture ($\Gamma_{c}$) configuration of the \emph{Z}-scan
technique is given by Eq. (1) \cite{Sheik,Chapple}. This is the case
for systems that exhibit both nonlinear refraction and nonlinear absorption,
with the assumptions of local cubic nonlinearity, sufficiently thin
sample and to first-order corrections in the irradiance of a Gaussian
laser beam at the far field condition.
\begin{equation}
\Gamma_{c}=1-\frac{4\Phi\left(\frac{z}{z_{o}}\right)}{\left[1+\left(\frac{z}{z_{o}}\right)^{2}\right]\left[9+\left(\frac{z}{z_{o}}\right)^{2}\right]}-\frac{\Theta\left[\left(3+\left(\frac{z}{z_{o}}\right)^{2}\right)\right]}{\left[1+\left(\frac{z}{z_{o}}\right)^{2}\right]\left[9+\left(\frac{z}{z_{o}}\right)^{2}\right]}.
\end{equation}
 where $z_{o}$ is the Rayleigh range of the beam, $\Phi=k\, n_{2}I_{o}L_{ef}$,
$k$ is the wave number, $L_{ef}=\left[1-\exp\left(-\alpha_{o}L\right)\right]/\alpha_{o}$
is the effective thickness of the sample, $I_{o}$ is the irradiance
at the beam waist of the laser, $\alpha_{o}$ is the linear optical
absorption and $\Theta$ is proportional to the nonlinear absorption.
However under incidence of a moderated-power cw laser beam, the nonlinearity
in a nematic liquid crystals is essentially from thermal origin due
to the temperature dependence of the order parameter \emph{S}. It
is said that the laser induce a Thermal Lens and the intensity of
this effect is proportional to the thermooptical coefficient $dn/dT$\cite{Carter},
i.e. the refraction index can be written as \emph{$n=n_{0}+(dn/dT)\,\Delta T$,
}where $\Delta T$ is the change in \textcolor{black}{temperature
\cite{Simoni}. Alt}hough the diffusion of heat leads to a phase variation
of the laser beam which does not match exactly its intensity spatial
profile, for samples with low absorption and low thermal conductivity,
it was shown \cite{Cuppo} that the Shake-Bahae\textasciiacute{}s
model for the Z-scan experiment, based in a purely local effect, gives
a good description of the transmittance. In this case, it is possible
to write that $dn/dT\propto$$n_{2}$, where $n_{2}$ and $dn/dT$
are the fitting parameters of the Sheik-Bahae model \cite{Sheik}
and the Thermal Lens model, respectively.

Our experimental setup uses a cw laser ($\lambda=532\, nm$, Ventus,
Laser Quantum) with power in the range of $3-40\, mW$. The beam waist
at focus was about $26\,{\mu}m$, and data acquisition was made \emph{via}
oscilloscope (Tektronix). The polarization of the laser beam $\left(\mathbf{E}\right)$
was set either parallel or perpendicular to the nematic director $\mathbf{\left(n\right)}$
induced by the surface treatment, therefore, reorientation of the
nematic director by an optical torque $\left(\mathbf{\mathbf{\mathbf{\mathrm{\boldsymbol{\Gamma}}_{\mathrm{opt}}}}}=\mathbf{D}\times\mathbf{E}\right)$
is not expected to occur.

\section{Experimental results and discussions}

Fig. 2 shows the typical closed-aperture Z-scan trace, obtained with
sample S2 at $T=25\,^{o}C$ (nematic phase) for $\mathbf{n\parallel E}$.
The error bars correspond to the standard error of the mean value
for at least ten measurements in each \emph{z}-position. As can be
seen, for $\mathbf{n\parallel E}$ the sample exhibits the valley-peak
configuration corresponding to a negative nonlinear refraction coefficient
$\left(n_{2||}<0\right)$. On the other hand, for $\mathbf{n\perp E}$
the samples exhibit positive nonlinear refraction $\left(n_{2\perp}>0\right)$
(figure not shown). \textcolor{black}{These results are consistent
with the laser heating of the sample. An encrease of temperature leads
to a wider angular distribuition of the long molecular axis around
the nematic director. So, the effective molecule seen by the laser
beam is shorter along the molecular axis and thicker in a perpendicular
direction, resulting in a diminuition of the extraordinary refraction
index and an increase of the ordinary refraction index, respectively.
}We have also checked the nonlinear optical response of the empty
glass cell with PVA coating: for the intensities of the laser beam
used in our experiment, the cell does not show any nonlinear optical
response. 

Fig. 3 shows the nonlinear refraction of the samples in the nematic
and in the isotropic phases for both relative orientations between
the nematic director and the polarization of the laser as a function
of the normalized temperature $T/T_{NI}$. For $\mathbf{n\parallel E}$
(filled symbols) the samples display a self-defocusing effect $\left(n_{2||}<0\right)$
and for $\mathbf{n\perp E}$ (empty symbols) the samples display a
self-focusing effect $\left(n_{2\perp}>0\right)$ in the full nematic
region. Approaching $T_{NI}$ from below, the modulus of $n_{2}$
diverges for both relative orientations between the polarization of
the optical beam and the nematic director. Fig 4 shows the nonlinear
birefringences $\triangle n_{2}=n_{2||}-n_{2\perp}$ of the samples
as a function of temperature. As observed, the magnitude of the nonlinear
birefringence increases as $T_{NI}$ is approached from below. Although
in the nematic phase far from $T_{NI}$ the birefringence of all the
samples are rather similar, close to $T_{NI}$ the higher the doping
of dye, the bigger is the birefringence. As the dye absorption is
dominant over the liquid crystal E7 at 532 nm, the nonlinear response
measured here comes mainly from the dyes, and that explains the nonlinear
birefringence behavior with the dye concentration. It was shown \cite{Vinicius}
that in the nematic phase and close to the clearing point the nonlinear
birefringence can be written as 

\begin{equation}
\Delta n_{2}\propto\left[1-\frac{T}{T^{\dagger}}\right]^{\beta-1},
\end{equation}

where $\beta$ is the effective critical exponent associated with
the order parameter and $T^{\dagger}$, the temperature of the virtual
second order transition seeing from below $T_{NI}$, represents the
absolute limit of the nematic phase on heating. For the homologous
series of cyano-biphenyls \emph{n}CB, with \emph{n}=5 to 8, $T^{\dagger}-T_{NI}\simeq0.2\,^{\circ}C$\cite{Chirtoc},
being of the order of the uncertainness of our measurement of temperature.
Figure 5 shows a ln-ln plot of $\left\vert \Delta n_{2}\right\vert $
as a function of the reduced temperature $1-\frac{T}{T^{\dagger}}$
for sample S3 supposing $T^{\dagger}=T_{NI}$. Similar plots were
obtained for the other samples. To perform the best fitting we proceeded
to vary $T^{\dagger}$ at regular steps in the temperature range $[T_{NI},T_{NI}+0.4\,^{\circ}C]$
for each sample and the weighted averages of the effective critical
exponents $\beta$ are summarized in Table 1. 

The value of $\beta$ is rather independent of the dye doping concentration,
but is a little higher for the pure E7 sample, nevertheless they also
support the tricritical hypothesis of the N-I phase transition. It
is important to emphasize that the exponents $\beta$ obtained for
the doped samples (S1, S2 and S3) are mainly associated to the dye
order parameter. From previous work \cite{Alliprandini}we observed
that the dye order parameter increases in the range of concentration
investigated here, so it is possible to conclude that the critical
exponent $\beta$ is not directly affected by changes in the molecular
order parameter.

\section{Conclusions}

To summarize, it was investigated the nonlinear optical properties
and the nonlinear refraction of the dye-doped E7 liquid crystal in
the nematic phase, close to the N-I transition under continuous 532-nm
wave excitation by the Z-scan technique. The nonlinear optical refraction
exhibit opposite character for the two geometrical configurations
of the nematic director relative to the polarization of the incident
beam, being $n_{2||}<0$ and $n_{2\perp}>0_{2}$, in all the nematic
phase. Approaching the clearing point from below, the nonlinear refraction
diverges. The effect of the dye is to amplify the behavior observed
in the undoped sample. Curiously, the amplification of the nonlinear
response is higher for $\mathbf{n\perp E}$ than for $\mathbf{n\parallel E}$.
The nonlinear birefringence is characterized by a critical exponent
that seams to confirm the tricritical hypothesis of the N-I phase
transition wich predicts a value 0.25. The tricritical character of
the N-I phase transition seams to be independent of the E7 liquid
crysrtal doping.

\textbf{Acknowledgments}

This work had the financial support of the Brazilian agencies CAPES,
CNPq, FAPESP, Secretaria de Ciência, Tecnologia e Ensino Superior
do Paraná and Fundação Araucária, and it was conducted as part of
the research program of the Instituto Nacional de Ciência e Tecnologia
de Fluidos Complexos (INCT-FCx).

\bigskip{}

\begin{center}
\textbf{Figure Captions} 
\par\end{center}

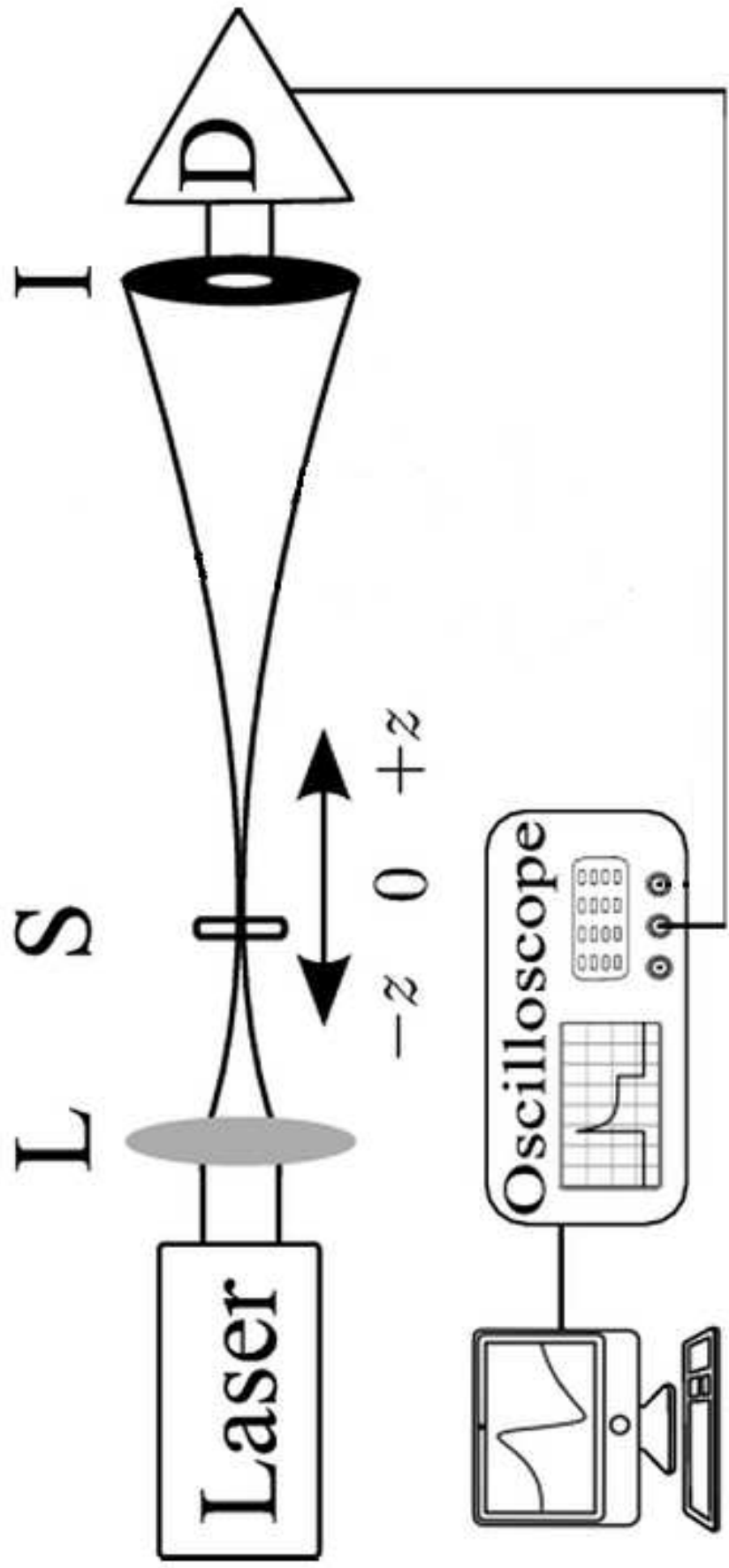
Fig. 1: Sketch of the apparatus for the implementation of the closed-aperture
Z-scan: L (lens), S (sample), I (iris), and PD (photodetector).\\

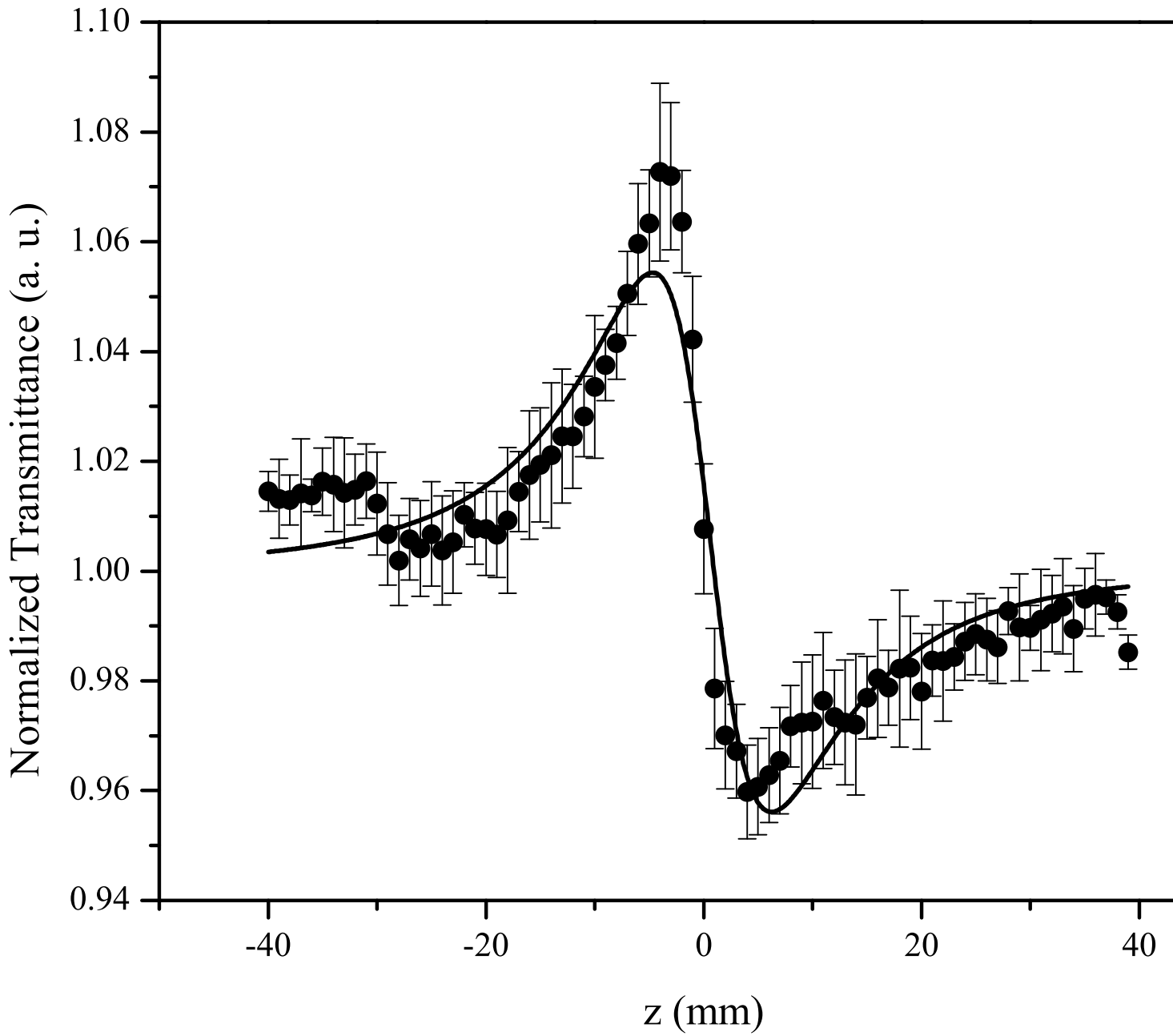
Fig. 2: Typical $Z$-scan curve obtained for $\mathbf{n}\parallel\mathbf{E}$
with closed-aperture configuration Z-scan technique ($T=25\,^{\circ}C$
). The solid line shows the fitting to Eq. 1.\\

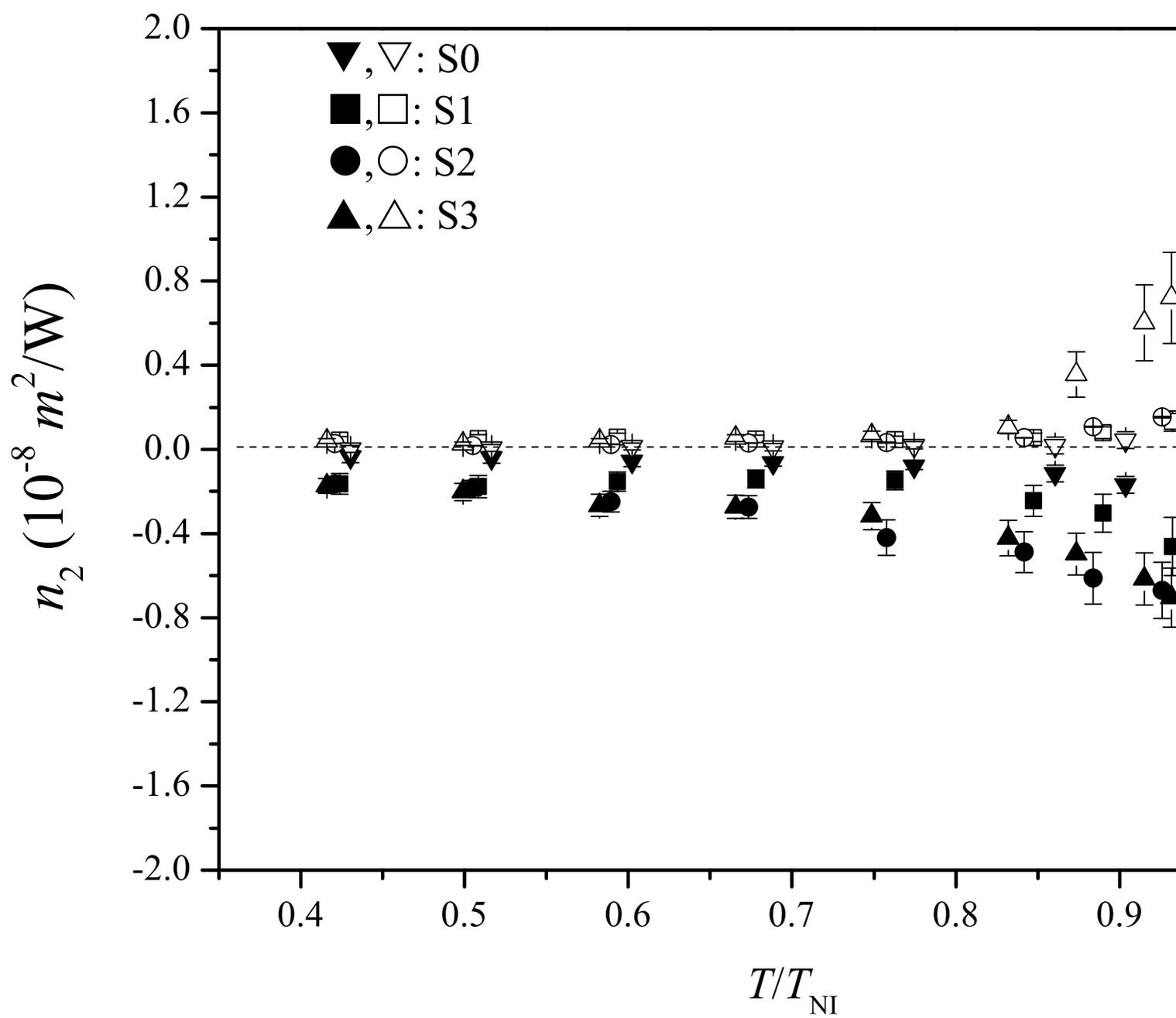
Fig 3: Nonlinear refraction as a function of temperature in the nematic
and isotropic phases, for both configurations between the nematic
director and the polarization of the incident beam: $\mathbf{n}\parallel\mathbf{E}$
(full symbols) and $\mathbf{n}\perp\mathbf{E}$ (empty symbols).\\

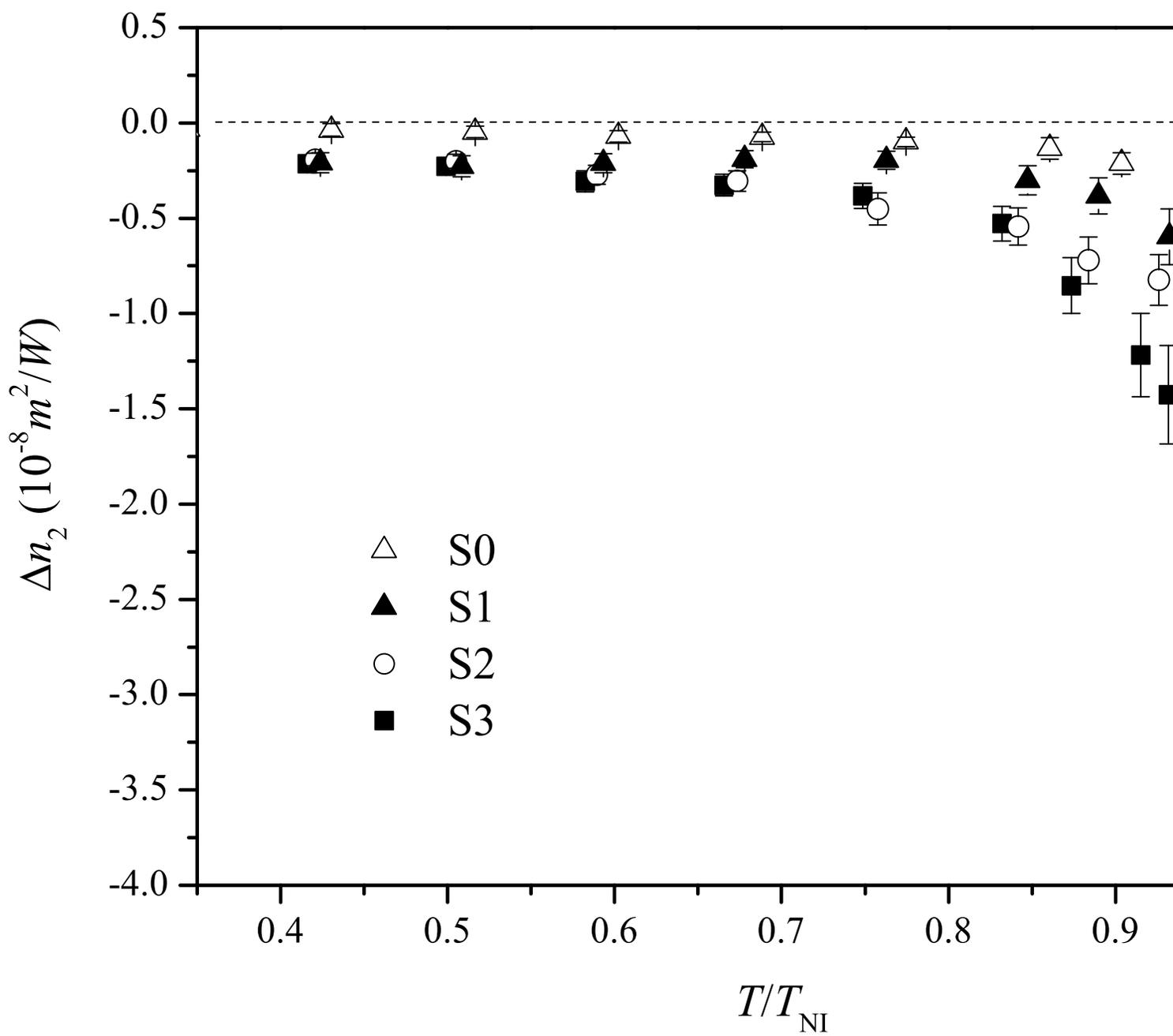
Fig. 4: Nonlinear birefringence $\triangle n_{2}$ of the samples
as a function of temperature. Vertical dashed line indicates $T_{NI}$.\\

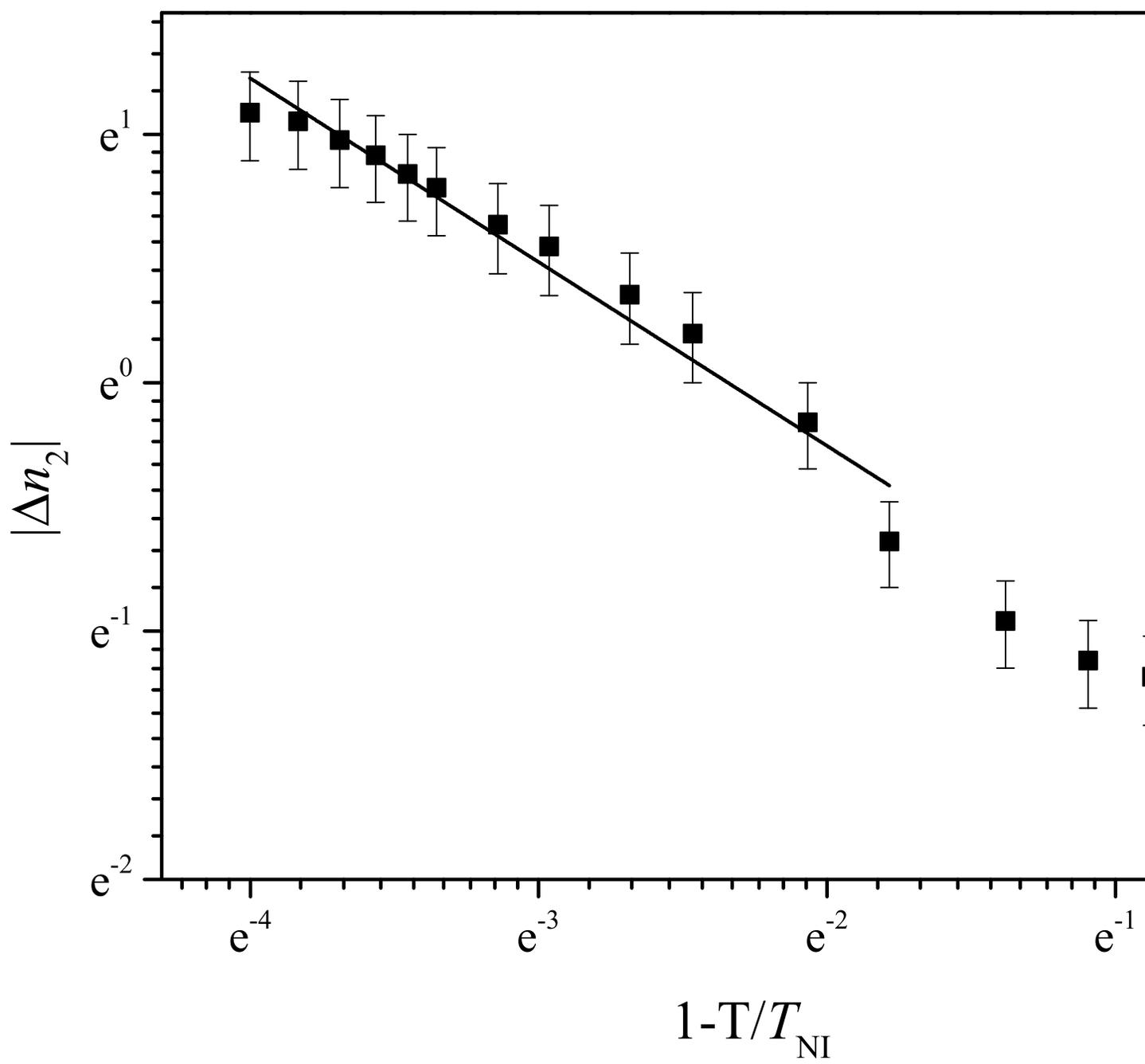
Fig. 5: ln-ln plot of the absolute value of $\Delta n_{2}$ as a function
of the reduced temperature $1-T/T_{NI}$. Solid line represents a
typical linear fitting of data.\\

\bigskip{}

\begin{center}
\textbf{Table Captions} 
\par\end{center}

Table 1: Values of the effective critical exponents $\beta$ associated
to the order parameter for the doped (S1, S2, S3) and the undoped
(S0) samples measured from the nonlinear birefringence $\triangle n_{2}$
at the N-I phase transition.\\

\bigskip{}
 \begin{center}
\textbf{Table} 
\par\end{center}

\begin{center}
\begin{tabular}{ccc} 

\hline\noalign{\smallskip}
Sample    && $\beta$  \\
\hline
S0       && 0.28 $\pm $ 0.03 \\
S1       && 0.22 $\pm $ 0.04 \\
S2       && 0.23 $\pm $ 0.03 \\
S3       && 0.21 $\pm $ 0.03 \\ 
\hline 
\end{tabular}
\par\end{center}
\end{document}